# Image denoising in acoustic field microscopy

Shubham Kumar Gupta[1†], Azeem Ahmad[2], Prakhar Kumar[3], Frank Melandsø[2], Anowarul Habib[2]

[1] Dept of Chemical Eng., Indian Institute of Technology Guwahati, India
[2] Dept. of Physics and Technology, UiT The Arctic University of Norway, Norway
[3] Dept of Electronics Eng, Indian Institute of Technology Dhanbad, India

## 1. Introduction

Scanning acoustic microscopy (SAM) has been employed since microscopic images are widely used for biomedical or materials research. Acoustic imaging is an important and well-established method used in nondestructive testing (NDT), bio-medical imaging, and structural health monitoring [1-3]. It provides abundant qualitative and quantitative information about the objects under inspection. Acoustic Microscopy utilizes low or high-frequency ultrasounds to produce visible images of the hidden regions of an object without damaging the object [4]. Furthermore, we can analyze and detect defects efficiently by thoroughly studying these objects' internal layers and structures.

SAM is a valuable tool in various production processes because it allows non-destructively viewing the internal structure of every sample and analyzing the potential impacts of the failure's root causes. This method efficiently detects physical defects, including fractures, voids, and delamination, with high sensitivity without harming the material under study by keeping track of a sample's internal features in three dimensions of integration. Artifacts or randomly occurring statistical changes are referred to as noise in the images.

Spatial domain and transform domain methods are two main categories in denoising techniques. In spatial filter domain is further categorized into linear and non-linear filters. It utilizes low pass filtering on image pixel values as the noise tends to occupy higher regions in the frequency spectrum. To denoise the images and extract the correct information from the images, a variety of techniques are available. The imaging is frequently carried out with signals of low amplitude, which might result in leading that are noisy and lacking in details of image information. In this work, we attempted to analyze SAM images acquired from low amplitude signals and employed a block matching filter over time domain signals to obtain a denoised image. We have compared the images with conventional filters applied over time domain signals, such as the gaussian filter, median filter, Wiener filter, and total variation filter. The noted outcomes are shown in this article.

---

email: anowarul.habib@uit.no

## 2. Experimental setup:

Data acquisitions were performed on a custom-built SAM (fig.1), integrated with a Standa (8MTF-200-Motorized XY Microscope Stage) high precision scanning stage controlled by LabVIEW [5]. Our group employed a similar experimental setup earlier to determine and correct the inclined sample [5]. The SAM features were implemented using National Instruments' PXIe FPGA modules and FlexRIO hardware. It was enclosed in a PXIe chassis (PXIe-1082), which consists of an arbitrary waveform generator (AT-1212).

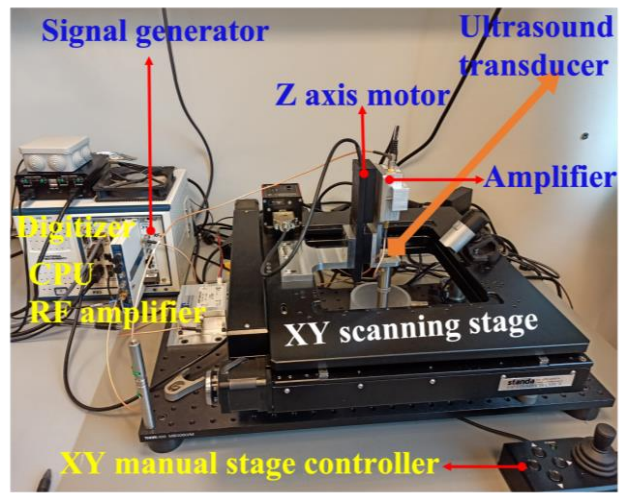

**Fig. 1:** The figure demonstrates the SAM used for image acquisition as discussed in this paper.

The excited signals were then amplified using a radio frequency (RF) amplifier (AMP018032-T) for amplification of the ultrasonic signals. For this experiment, an Olympus 50 MHz focused transducer with an aperture of 6.35 mm and a focal length of 12 mm was used to scan the coin samples.

## 3. Results and discussions

The volumetric data of a 10-cent coin was taken, and ultrasonic time signals of the data were subjected to several denoising filters, including block matching and Wiener filters. The block matching algorithm is one of the strongest and most efficient volumetric data signal denoising methods. The algorithm uses a higher dimensional transform-domain representation to enforce sparsity and hence regularize the data. This is termed nonlocal grouping and collaborative filtering. The suggested approach takes advantage of the 3-D spatiotemporal volumes created by monitoring blocks along motion vector-defined trajectories' mutual resemblance[6]. To achieve collaborative filtering, each group is transformed using a decorrelating 4-D separable transform, followed by shrinkage and inverse transformation[7-9]. Collaborative filtering generates estimates for each volume stacked in the group, which are then adaptively aggregated back to their original place.

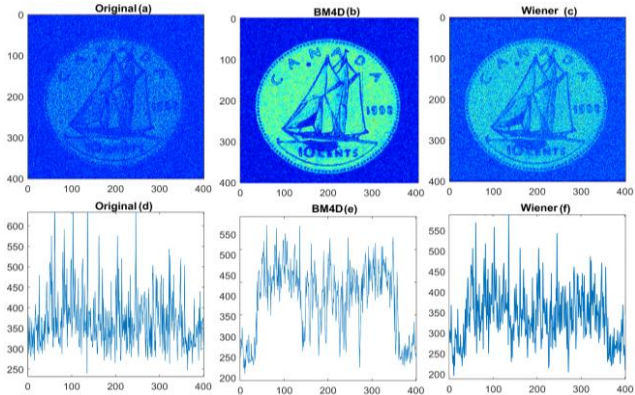

**Fig 2**: Image comparison with the different filters applied in the paper. The lower figure (d, e, f) represents the line profile of the corresponding image (a, b, c).

Here figure 2, (a) shows the original image, i.e., 0.21V amplitude image, (b) Wiener filtered image, and (c) 4D block matching filter image applied on the time domain signals. The second row shows the line profile of the corresponding image at the center. The line profile demonstrates the noises present in the original image.

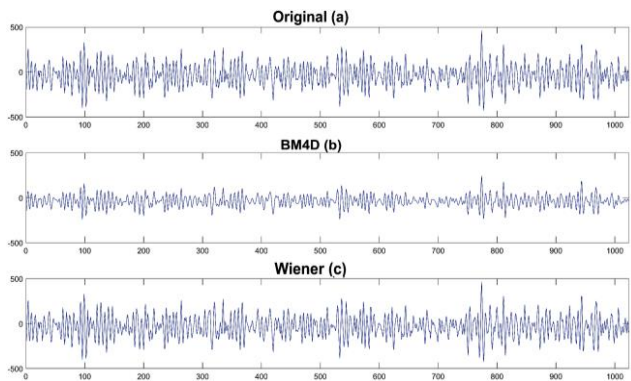

**Fig. 3:** Represents the time domain signals of the original, weiner filtered, and 4D block matching filtered data.

Here figure 3 shows the various time domain signals filtered using various filters like (a) shows the original data time signals, scanned at an amplitude of 0.21V, (b) shows the 4D-block matching filtered time signals, and (c) shows the Wiener filtered time signals.

## 4. Conclusion

In this paper, we have demonstrated a 4D block matching filter that can be used to denoise acoustic volumetric signals. We have compared it with a conventional denoising Wiener filter and compared the image with our proposed block matching filer. From the visual inspection of the image, it is evident that the proposed block match filter performed better than the conventional Wiener filter. The proposed block matching filter would be a good option in image denoising where the signal-to-noise ratio is poor, like in photoacoustic imaging.

**Acknowledgment**

UiT funded Tematisk Satsinger project VirtualStain (Cristin Project ID: 2061348)